\documentclass[conference]{IEEEtran}
\usepackage{amssymb}
\usepackage{amsmath}

\usepackage[dvips]{graphicx}
\usepackage{epsfig}
\usepackage[font=footnotesize]{subcaption}
\usepackage{longtable}
\usepackage{booktabs}

  \usepackage[font=footnotesize]{subfig}

\ifCLASSINFOpdf
\else
\fi

\begin{document}
%
\title{Evolutionary stable strategies in networked games: the influence of topology}

\author{\IEEEauthorblockN{Dharshana Kasthurirathna}
\IEEEauthorblockA{Centre for Complex Systems Research\\
Faculty of Engineering and IT\\
The University of Sydney, NSW 2006\\
Australia\\
Email: dkas2394@uni.sydney.edu.au}
\and
\IEEEauthorblockN{Mahendra Piraveenan}
\IEEEauthorblockA{Centre for Complex Systems Research\\
Faculty of Engineering and IT\\
The University of Sydney\\
NSW 2006\\
Australia\\ 
Email: mahendrarajah.piraveenan@sydney.edu.au}
\and 
\IEEEauthorblockN{Shahadat Uddin}
\IEEEauthorblockA{Centre for Complex Systems Research\\
	Faculty of Engineering and IT\\
	The University of Sydney\\
	NSW 2006\\
	Australia\\ 
	Email: shahadat.uddin@sydney.edu.au}}


%


\maketitle

\begin{abstract}
Evolutionary game theory is used to model the evolution of competing strategies in a population of players. Evolutionary stability of a strategy is a dynamic equilibrium, in which any competing mutated strategy would be wiped out from a population. If a strategy is weak evolutionarily stable, the competing strategy may manage to survive within the network. Understanding the network-related factors that affect the evolutionary stability of a strategy would be critical in making accurate predictions about the behaviour of a strategy in a real-world strategic decision making environment. In this work, we evaluate the effect of network topology on the evolutionary stability of a strategy. We focus on two well-known strategies known as the Zero-determinant strategy and the Pavlov strategy. Zero-determinant strategies have been shown to be evolutionarily unstable in a well-mixed population of players. We identify that the Zero-determinant strategy may survive, and may even dominate in a population of players connected through a non-homogeneous network. We introduce the concept of `topological stability' to denote this phenomenon. We argue that not only the network topology, but also the evolutionary process applied and the initial distribution of strategies are critical in determining the evolutionary stability of strategies. Further, we observe that topological stability could affect other well-known strategies as well, such as the general cooperator strategy and the cooperator strategy. Our observations suggest that the variation of evolutionary stability due to topological stability of strategies may be more prevalent in the social context of strategic evolution, in comparison to the biological context. 
\end{abstract}


%
\IEEEpeerreviewmaketitle

\section{Introduction}
Game theory is the science of strategic decision making among autonomous players \cite{von1944game}. Game theory has its roots in micro-economics and has later been adopted in myriad fields of study, such as biology, psychology and computer science \cite{rasmusen1994games}. Evolutionary game theory is the branch of study that has resulted from the adoption of game theory into evolutionary biology \cite{smith1973lhe}. It is used to study how a particular strategy or a group of strategies would evolve over time in a population of players. If a strategy is an evolutionarily stable strategy (ESS), once it is adopted by a population of players, any mutated strategy would not be able to invade it \cite{smith1993evolution}. Evolutionary stability of a strategy could further be divided into two sub categories. strong ESS and weak ESS (also called asymptotic stable strategy and stable strategy \cite{bendor1995types}). If a strategy is in a weak evolutionarily stable state, the invading strategy does not completely die-out but its population does not increase \cite{bendor1995types}. 

Meanwhile, complex  networks science is increasingly used to model and analyse human interactions~\cite{barabasi1999emergence,statmech2002}.  Many societies where humans or other sentient agents  interact display heterogeneous network topologies, in which some agents / people are very well connected while others are sparsely connected~\cite{statmech2002,thedchanamoorthy2014influence}. In such societies or multi-agent networks~\cite{prokopenko2005convergence}, the topology of the  network of interactions can be quantified by a range of metrics, including the clustering coefficient, average path length, global and local assortativity~\cite{Newman2003,EPL,PPZ}, topological  mutual information~\cite{sole,piraveenan2007emergence}, various static and dynamic centrality measures~\cite{piraveenan2013percolation} and so forth. Moreover, many of these social systems display the so-called scale-free property, which is characterised by maximum heterogeneity in terms of number of connections, and the small-world property, where the characteristic path length is comparatively small and the clustering comparatively high. Therefore it is clear that the topology of any society or agent network must be considered in determining the dynamics of a system, including how individual agents make decisions in such systems.

In this work, we study how network topology affects the evolutionary stability of strategies. We use a class of strategies known as `memory-one strategies' in prisoner's dilemma (PD) game to evaluate the effect of network topology on evolutionary stability. Iterated prisoner's dilemma game has widely been used to model -the strategic decision making of self-interested opponents \cite{santos2005scale, fogel1993evolving, adami2013evolutionary}. In memory-one strategies, each player would base his action on a probability derived based on the previous interaction with the same opponent~\cite{kasthurirathna2014influence,kasthurirathne2013optimization}. In this work, we particularly focus on a sub-class of memory-one strategies known as `Zero-Determinant(ZD) strategies', along with other well-known memory-one strategies. Zero-determinant strategies have been demonstrated to be extortionate strategies, meaning that they have the ability to unilaterally set the payoff of the opponent \cite{press2012iterated}. Intuitively, this would suggest that Zero-Determinant strategies have the potential to be evolutionarily stable against any competing strategy. However, Zero-determinant strategies have later been shown to be evolutionarily unstable in a well-mixed population of players \cite{adami2013evolutionary}. In this work, we test whether network topology affects the evolutionary stability of Zero-determinant strategies in a non-homogeneous network of players.

To test the effect of network topology, we use two well known network classes, known as the scale-free networks \cite{statmech2002} and well-mixed networks. Also, we use two evolutionary processes to evolve the populations. First, we test the effect of topology using the death-birth Moran process \cite{moran1962statistical}, which is an evolutionary process used to model the evolution of players over time, particularly in biological systems. Then we use a stochastic strategy adoption process that would update the strategy of a randomly selected node, by comparing it with a selected neighbour's strategy. The same process has been used as an evolutionary process by Santos et al. \cite{santos2006graph}, with the pure-strategy PD game. We will refer to this evolutionary process as the `strategy adoption process' in the rest of the paper. We compare and contrast the evolutionary outcomes of these two processes when players are placed in both well-mixed networks and scale-free networks. Based on the results of these observations, we argue that the evolutionarily unstable strategies in a well-mixed population may survive and even dominate in a heterogeneous network of players. We show that the topology of the interactions of the players, the evolutionary update process as well as the initial topological distribution of players are significant determining the overall evolutionary stability of a strategy. When the players are distributed in a homogeneous network however, the evolutionary process used would not have a significant effect the evolutionary stability of a strategy. We further test the effect of topology on the evolutionary stability by varying network assortativity \cite{Newman2003}, which is a measure of the similarity of mixing of nodes in a network. We argue that when the network becomes more heterogeneous, network topology would have a more significant effect on the evolutionary stability of strategies. We call this topologically influenced evolutionary stability of strategies as `topological stability'.

Understanding the topological effect on the evolutionary stability of a strategy would help us to make better predictions about the evolutionary stability of a strategy in a real-world environment. Even though a strategy may be theoretically stable or not, its actual evolutionary behaviour may depend on the topology of the interconnections in the population and the evolutionary update mechanism used. By studying these effects extensively, the modelling of evolutionary games may be improved, by increasing the accuracy of the predictions of the evolutionary stability of strategies.

The rest of the paper is organised as follows. We begin by providing a background on the theoretical aspects of evolutionary game theory and complex network science, within the scope of this work. Therein we introduce the memory-one strategies in the iterated prisoner's dilemma (IPD) game, as well as a sub-class of memory-one strategies known as Zero-determinant strategies. Further, we explain death-birth Moran process and its significance in determining the evolutionary stability of a strategy, comparing it with the strategy adoption process suggested by Santos et al. \cite{santos2006graph}. Next, we present the results obtained by simulating both the death-birth Moran process and the strategy adoption process in well-mixed and scale-free networks of players. Further, we present the results on how the variation of network heterogeneity, measured using the network assortativity, affects the evolutionary stability of strategies. Finally, we discuss the results, presenting our conclusions. 

\section{Background}

\subsection{Game Theory}

Game theory is the study of strategic decision making \cite{von1944game}. Game theory was first developed as a branch of micro-economics \cite{von1944game, rasmusen1994games}. However, later it has been adopted in diverse fields of study, such as evolutionary biology, sociology, psychology, political science and computer science \cite{rasmusen1994games}. Game theory has gained such wide applicability due to the prevalence of strategic decision making across different fields of study. One of the key concepts of Game theory is that of Nash equilibrium \cite{nash1950equilibrium}. Nash equilibrium suggests that there is an equilibrium state in a strategic game, which neither player would benefit deviating from. Nash equilibrium can be defined for both pure strategy and mixed strategy scenarios. Also, a strategy game could have multiple Nash equilibria. 

A formal definition for Nash equilibrium can be given as follows. Let $(S, f)$ be a game with $n$ players, where $S_i$ is the strategy set of a given player $i$. Thus, the \textit{strategy profile} $S$ consisting of the strategy sets of all players would be, $S = S_1 \times S_2 \times S_3.... \times S_n$. $f = (f_1(x),.....,f_n(x))$ would be the payoff function for $x \epsilon S$. Suppose $x_i$ is th strategy profile of player $i$ and $x_{-i}$ be the strategy profile of all players except player $i$. Thus, when each player $i \epsilon {1,.....,n}$ chooses strategy $x_i$ that would result in the strategy profile $x= (x_1,....,x_n)$, giving a payoff of $f(i)$ to that particular player. A strategy profile $x^{*} \epsilon S$ is in Nash equilibrium if no unilateral deviation in strategy by any single player would return a higher utility for that particular player. Formally put,

\begin{equation}
\forall i,x_i \epsilon S_i : f_i(x_i^*, x_{-i}^*) \geq f_i(x_i,x_{-i}^*)\label{eq_nash}
\end{equation}

\subsection{Evolutionary Game Theory}

Evolutionary game theory is an outcome of the adaptation of game theory into the field of evolutionary biology \cite{smith1973lhe, smith1993evolution}. Evolutionary biology is based on the idea that an organism's genes largely contribute to its observable characteristics. Thus, genes determine the fitness of an organism in a given environment. Evolutionary game theory argues that the success of an organism depends on its interaction with other organisms within a population. Thus, the fitness of an organism would be dependent on its interaction with other organisms in that population. In game theoretic terminology, the interactions of players would be the strategies and the fitness of each strategy would be the payoff. As the population evolves over time, certain strategies would be dominant while some other strategies would be extinct. Thus, evolutionary game theory provides a mathematical basis to model the evolutionary process using concepts of game theory. On the other hand, it opens a new dimension for game theorists to observe how the evolution of strategies in games would happen over time in populations of players. Some of the critical questions asked in EGT include; which populations/strategies are stable? When to individuals adopt other strategies? and would it be possible for mutants to invade a given population?

 The equivalent concept to Nash Equilibrium in evolutionary game theory, is evolutionary stability \cite{maynard1974theory}. If Nash Equilibrium can be considered as a static equilibrium, evolutionary stability represents a dynamic equilibrium of a strategy, over time. A strategy is called evolutionarily stable if it has the potential to dominate over any mutant strategy \cite{smith1973lhe}. Evolutionary games are often modelled as iterative games where a population of players play the same game iteratively in a well-mixed or a spatially distributed environment \cite{le2007evolutionary}.

\subsubsection{Iterated prisoner's dilemma}

\begin{figure}[htbp]
\begin{center}
\resizebox{0.5\textwidth}{!}{
\includegraphics[width=7.0cm, height=5.0cm]{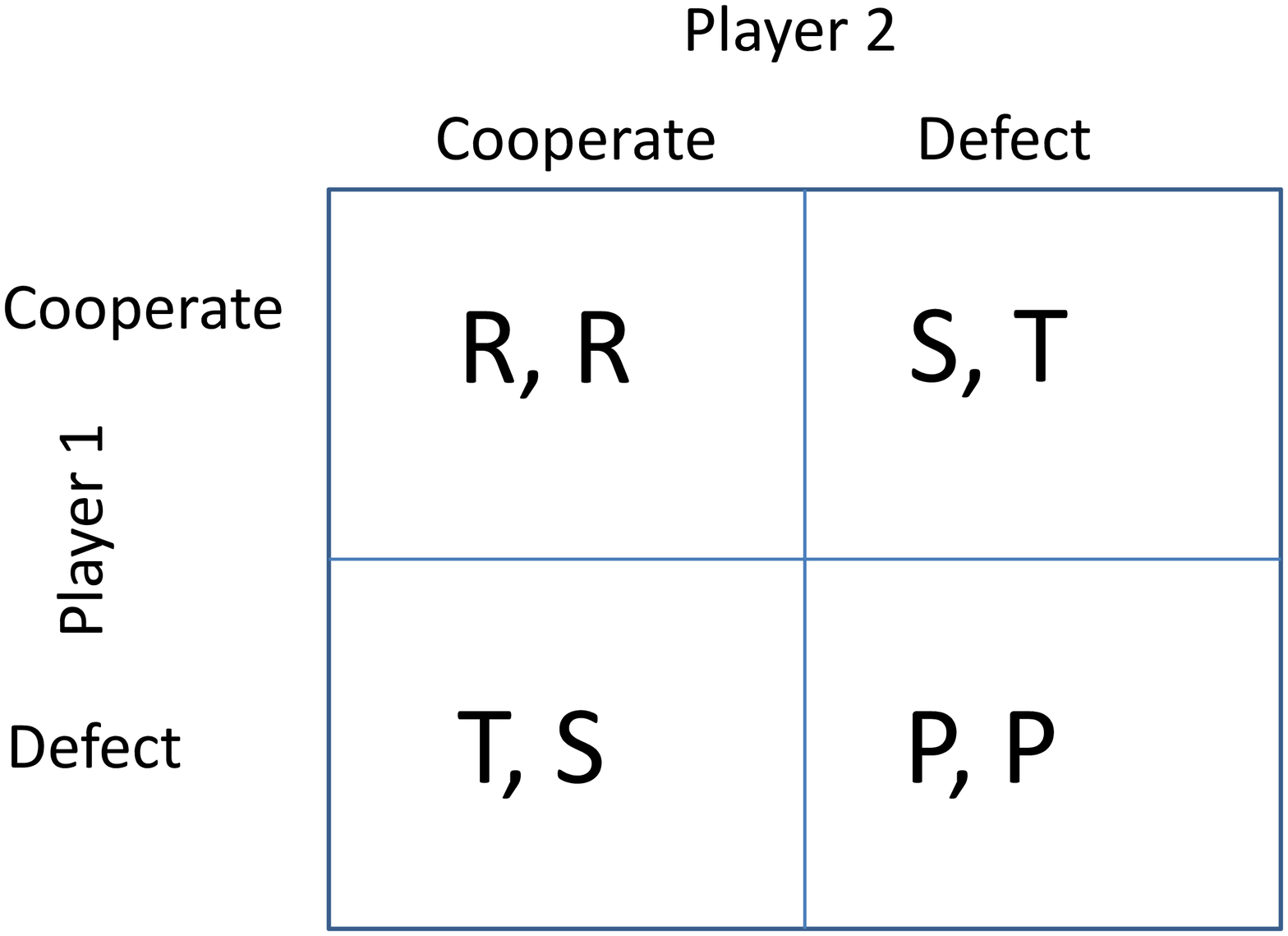} 
}
\caption{The payoff matrix of the prisoner's dilemma game. T\textgreater R\textgreater P\textgreater S} \label{f0}
\end{center}
\end{figure}

Prisoner's dilemma is a game that is found in classical game theory \cite{rapoport1965prisoner}. Given the payoff matrix in Fig. \ref{f0}, the inequality T\textgreater R\textgreater P\textgreater S should be satisfied in a prisoner's dilemma game. In other words, in the prisoner's dilemma game, the highest mutual payoffs are obtained by the players when both players cooperate. However, if one player cooperates while the other defects, the defector would obtain a higher payoff. The dilemma is that the Nash Equilibrium of this game, which occurs when both players defect, does not provide the optimum payoff for both the players.

In iterated prisoner's dilemma(IPD), the prisoner's dilemma game is iterated over  many time-steps, over a population of players \cite{fogel1993evolving}. Each player would play a single iteration of the game with its neighbours in each time-step. Iterated prisoner's dilemma game is widely used to model the autonomous decision making behaviour of self-interested players. It has been demonstrated that the topology of the network is significant in the evolution of cooperation of strategies in the IPD game \cite{santos2006graph}. For example, when the iterated prisoner's dilemma game is played among pure cooperation and pure defection strategies, cooperation evolves to be the dominant strategy in a population of players that are distributed in a scale-free topology.



\subsubsection{Memory-one strategies}

As opposed to pure strategies of cooperation and defection, mixed strategies of prisoner's dilemma game are based on the assumption that each player chooses a strategy based on a probability distribution. In fact, pure strategies can be regarded as a special case of mixed strategies where each strategy is chosen with the probability of one. Memory-one strategies \cite{adami2013evolutionary, kraines2000natural} are a special sub-class of mixed strategies, where the current mixed strategy of a game would depend on the immediate previous interaction between the two players in concern. On the other hand, Memory-one strategies are a specialisation of a more general class of strategies called finite-memory strategies \cite{adami2013evolutionary, kraines2000natural}, where the current mixed strategy would be dependent on $n$ number of historical states between the two players.  

When considering the previous state between two players in a PD game, there could be four possible states. Namely CC, CD, DC and DD, where C represents cooperation and D represents defection, respectively. Memory-one strategies are represented by calculating the probabilities of cooperation by a player in the next move, given the type of the previous interaction of the player with the same opponent. For example, a strategy (1,1,1,1) would imply that the Player A would cooperate with player B, irrespective of the previous encounter between Player A and B. Similarly, (0,0,0,0) would represent a player that always defects. Thus, the pure strategy cooperation and defection can be thought of as a special case of memory-one or finite memory strategies. By varying the probabilities of cooperation under each of the previous encounters, it is possible to define any number of mixed strategies. Some of the well-known memory-one strategies include the Pavlov strategy (1,0,0,1) and the general cooperator (0.935, 0.229, 0.266, 0.42) strategy. General cooperator is the evolutionarily dominating strategy that evolved at low mutation rates as demonstrated by Iliopoulos et al. \cite{iliopoulos2010critical}. Tit-for-Tat, is another well knowns strategy where a player would only cooperate if the opponent cooperated in the previous interaction. This is signified by the probability distribution (1,1,0,0).

\subsubsection{Zero-Determinant strategies}

Zero-determinant strategies \cite{press2012iterated, stewart2012extortion} are a special sub-class of memory-one strategies that have recently gained much attention in the literature and media. ZD strategies denote a class of memory-one strategies that enable a player to unilaterally set the opponent's payoff. Due to this inherent property, ZD strategies have the ability to gain a higher expected payoff against an opposing strategy. However, for a strategy to be evolutionarily stable, it has to be stable against itself as well as the opponent strategies. It has been shown that ZD strategies do not perform well against itself.  Due to this reason, ZD strategies have been demonstrated to be evolutionarily unstable \cite{adami2013evolutionary}, particularly against the Pavlov strategy.


ZD strategies are defined using a set of conditional probability equations \cite{press2012iterated}. Suppose $p_1$, $p_2$, $p_3$ and $p_4$ denote the set of probabilities that a player would cooperate given that the player's last interaction with the same opponent resulted in the outcomes CC ($p_1$), CD ($p_2$), DC ($p_3$) or DD ($p_4$). ZD strategies are defined by fixing p2 and p3 to be functions of $p_1$ and $p_4$, denoted by Eq. \ref{eq1} and Eq. \ref{eq2}.

\begin{equation}  \label{eq1}
p_2 = \frac{p_1 (T -P) - (1 + p_4)(T - R) }{R - P} 
\end{equation}

\begin{equation}  \label{eq2}
p_3 = \frac{(1-p_1)(P - S) + p_4(R - S) }{R - P} 
\end{equation}

It was shown by Press and Dyson \cite{press2012iterated} that when playing against the ZD strategy, the expected utility of an opponent O can be defined using the probabilities $p_1$ and $p_4$, while $p_2$ and $p_3$ are defined as functions of $p_1$ and $p_4$. Eq. \ref{eq3} gives the expected payoff of the opponent against the ZD strategy.

\begin{equation}  \label{eq3}
E(O, ZD) = \frac{(1-p_1)P + p_4R)}{(1-p_1+p_4)}
\end{equation}

Here, P and R represent the payoffs earned when both players defect and corporate, respectively.

Hence, ZD strategies allow a player to unilaterally set the opponent's payoff, effectively making them extortionate strategies. In the simulations performed here, we set the probabilities $p_1$ and $p_4$ as 0.99 and 0.01 respectively, as in the study done by Adami and Hintze \cite{adami2013evolutionary}. We then derive $p_2$ and $p_3$ to be 0.97 and 0.02, using the ZD conditional probability equations Eq. \ref{eq1} and Eq. \ref{eq2}.










\subsection{Evolutionary processes}

We make use of two evolutionary processes in the evolution of populations. The first one is a well-known evolutionary process known as the death-birth Moran process. The second one is the stochastic strategy adoption process that was adopted from the work of Santos et al. \cite{santos2006graph}. 

\subsubsection{Death-birth Moran Process \cite{moran1962statistical}}

As the name suggests, in the death-birth Moran process, a node is randomly selected for removal at each time-step. Then, its replacement node is selected from its neighbours based on a probability proportional to the fitness of the neighbours. In the case of the iterated prisoner's dilemma game, the fitness is equivalent to the accumulated payoff of each node, averaged over its number of neighbours. Then, the selected neighbour is replicated to replace the node that is being removed. The new node would have zero payoff yet it would still have the same neighbours as the previous node that existed in the same topological space. This process is continued over $n$ number of time-steps to evolve the entire population over time. Death-birth Moran process is commonly used to emulate the evolution of biological species where the strategies are `hard-wired' into the players. If the lifetime of a player is significantly less than the time-span of evolution, as with the case of biological evolution, death-birth Moran process may effectively be used to simulate the evolution of strategies(players) over time. 

\subsubsection{Stochastic strategy adoption process}

Since Moran process maybe be more applicable in the biological context where the players with hard-wired strategies get replaced, it does not take into account the individual payoff differences of the node being replaced and the replicating node. Thus, it maybe possible that the node being replaced would actually have a higher cumulative payoff (fitness) than the replicating node. On the other hand, in the social context, the time-span of evolution of strategies could be considerably less than the lifetime of a player. Hence, players would be more inclined to adopt the apparently successful strategy and survive without getting replaced from the population.  In order to model this kind of social evolution, a stochastic strategy adoption process can be applied. Such a process has been used in Santos et al. \cite{santos2006graph} to demonstrate the evolution of cooperation in IPD games with pure strategies. We extend that method for mixed strategies in this work. When employing this particular strategy adoption process, a node going through evolution is not directly replaced. Instead, its strategy could be updated by comparing its cumulative payoff with that of a stochastically selected neighbour. As with the Moran process, in each time-step, a node is marked for update. Then, a potential node to compare it with is selected from its neighbours, based on a probability proportional to the fitness (accumulated payoff) of the neighbours. Then, the probability of the marked node adopting the strategy of its selected neighbour is calculated using the following equation.

\begin{equation}
p={\max  \{0,\ (P_y-\ P_x)/[k_>\ }(E\left(ZD,Pav\right)-\ E\left(Pav,ZD\right)]\} 
\end{equation}

Where, \\
$p$ - Probability that node \textit{X} would adopt \textit{Y}'s strategy \\
$P_x$ - Cumulative pay off of node \textit{X}\\
$P_y$ - Cumulative pay off of node \textit{Y}\\
$k_>$ - Maximum degree of \textit{X}'s degree (k${}_{x}$) and \textit{Y}'s degree (k${}_{y}$)\\
$E\left(ZD,Pav\right)-\ $ The expected payoff of a ZD node against a Pavlov node\\
$E\left(Pav,ZD\right)$-- The expected payoff of a Pavlov node against a ZD node\\

As shown above, the population update probability depends on the payoff difference of the marked node and the selected neighbour node. Also, the degree of those two nodes are used to normalise the effect of degree differences. Still, the cumulative payoff of the node with the higher degree would be higher due to the fact that it would have more interactions with other players. Thus, this equation implicitly captures the network topology in calculating the adoption probability. Due to this reason, this particular strategy adoption process can be used to study the topological effect on the evolutionary stability of strategies.

\subsection{Complex Networks}

Complex Networks are self-organising networks that show non-trivial topological features \cite{statmech2002}. Complex Network analysis provides a network perspective in analysing complex systems. Different classes of complex networks have been defined to model real-world complex systems such as social and biological systems. In this work, we mainly focus on two such network classes known as well-mixed networks and scale-free networks that are widely discussed in the network analysis literature. 

\subsubsection{Well-mixed Networks}

Well-mixed networks has been one of the most widely used network models in network analysis since the inception of complex network science. Also, they are widely used in modelling evolutionary games \cite{iliopoulos2010critical}. These networks are modelled, assuming that each node is connected to every other node in the network. Due to this reason, well-mixed networks can be regarded as a class of homogeneous networks. Therefore, it is possible to approximate well-mixed networks to lattice networks with each node having the identical number of neighbours. However, most of the real-world networks such as social networks and biological networks do not demonstrate well-mixed behaviour, instead being spatially distributed with non-homogeneous topological features \cite{statmech2002}. Still, well-mixed networks provide a good reference model to compare with the non-homogeneous networks \cite{santos2005scale}.

\subsubsection{Scale-free Networks}

Scale-free networks demonstrate a network topology that shows a power-law degree distribution \cite{barabasi1999emergence}. In other words, the degree distribution of the network would fit in a equation of the form y = $\alpha$x$^{-\gamma}$. Here, $\gamma$ is called the scale-free exponent. The scale-free exponent is obtained by fitting a particular degree distribution into a power-law curve. As the scale-free exponent increases, so does the power-law nature of the degree distribution. Also, if the correlation of the fitting is higher, such a degree distribution would indicate that the respective network closely resembles a scale-free network.

Scale-free nature is abundant in real-world networks, such as in social, biological and collaboration networks \cite{statmech2002}. Scale-free networks make perfect candidates to study the topological effect of a population of players due to this reason. For example, it has been shown that cooperation becomes the dominant strategy in a scale-free topology due to the heterogeneity of the network \cite{santos2005scale, santos2006graph}. Moreover, scale-free networks can be conveniently generated using the Preferential attachment growth model proposed by Barabasi and Albert \cite{statmech2002}. Preferential attachment model suggests that when a network grows in size, nodes with higher degree have a higher probability of attracting new nodes. In other words, there exists a degree-dependent `preference' in creating links with the existing nodes, when a new node joins the network. In this work, we use scale-free networks generated using the preferential attachment model for all our simulations.

\subsubsection{Network Assortativity}

Assortativity is the tendency observed in networks where nodes mostly
connect with similar nodes. Typically, this similarity is interpreted in terms
of the degrees of nodes. Assortativity has been formally defined as a correlation
function of excess degree distributions and link distribution of a network \cite{Newman2003, sole}.  The formal definition of Assortativity makes use of network topological concepts such as degree distribution $p(k)$ and excess degree distribution $q(k)$ for undirected networks \cite{sole}. Given $q(k)$, one can introduce the quantity $e_{j,k}$ as the joint probability distribution of the remaining degrees of the two nodes at either end of a randomly chosen link. Given these distributions,
the assortativity of an undirected network is defined as:

\begin{equation}
\label{eq1.8}\rho=\frac{1}{\sigma _q^2 }\left[ {
\sum\limits_{jk} {jk\left( {e_{j,k} -q_j q_ k} \right)} } \right]
\end{equation}

where $\sigma_q$ is the standard deviation of $q(k)$. In this work, we use networks with varying assortativity values to observe how the heterogeneity of mixing patterns affect the evolutionary stability of strategies. 




\section{Methodology}

\begin{figure*}[htbp]
\centering
	\begin{subfigure}{0.5\textwidth}
		\centering
		\includegraphics[width=7.0cm, height=5.0cm]{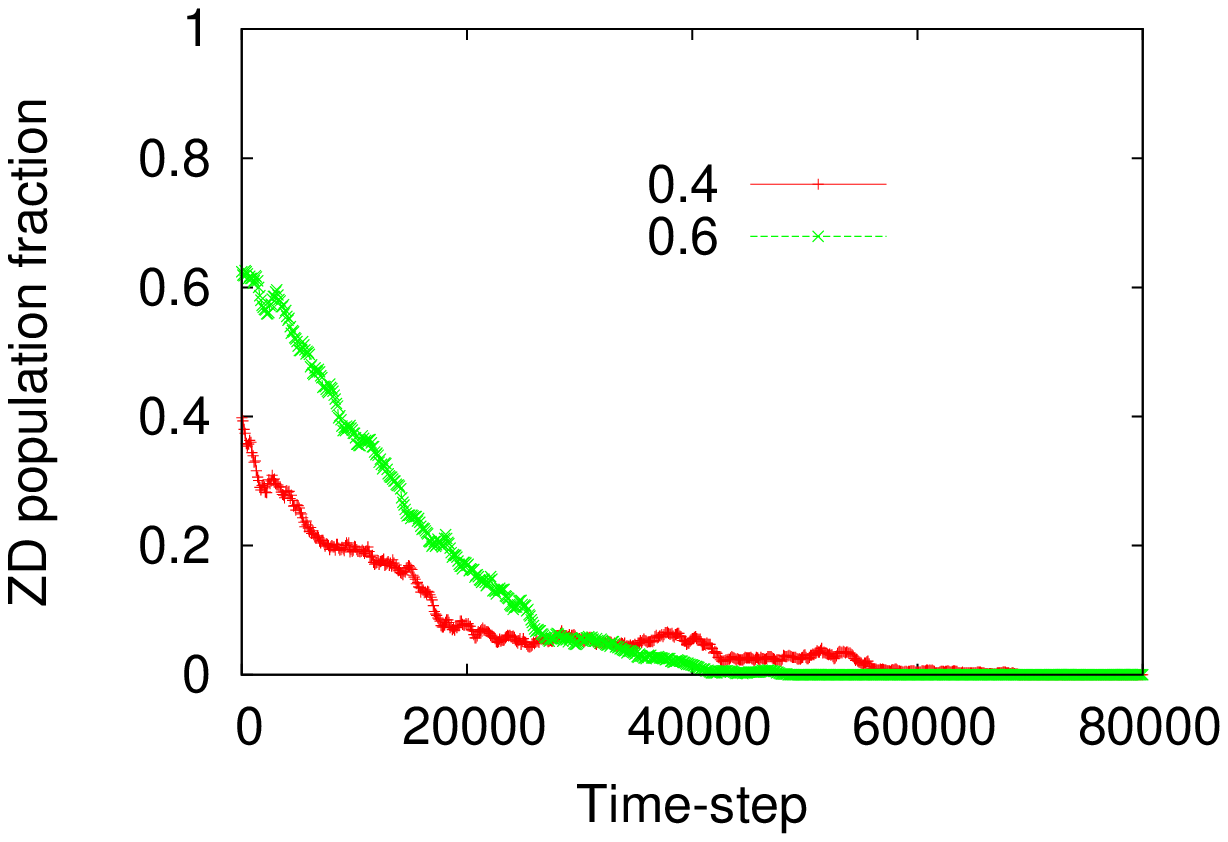}
		\caption{$Well-mixed$} 
	\end{subfigure}~
	\begin{subfigure}{0.5\textwidth}
		\centering
		\includegraphics[width=7.0cm, height=5.0cm]{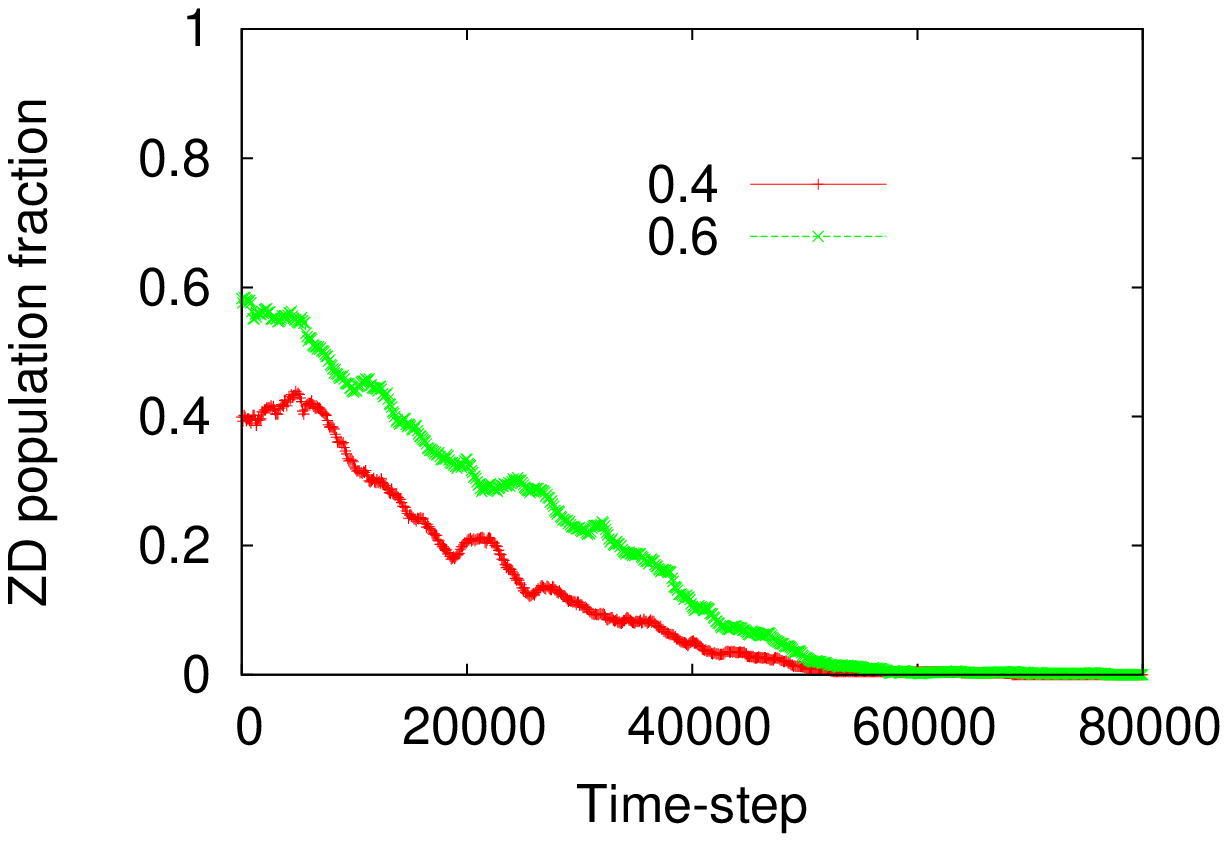}
		\caption{$Scale-free$} 
	\end{subfigure}
\caption{The evolution of ZD population fraction against the Pavlov strategy, in well-mixed and scale-free populations. The population is allowed to evolve under the death-birth Moran process with 0.1\% replacement rate. The strategies are initially distributed randomly, with the fraction of ZD nodes being 0.4 and 0.6, respectively.} \label{f1}  
\end{figure*}

Initially, we re-created the experimental results obtained in the work by Adami and Hintze \cite{adami2013evolutionary}, by mixing the Zero-determinant strategy with the Pavlov strategy in a well-mixed population. This further enabled us to confirm the theoretical results presented in the same work using the replicator dynamics model, suggesting that the ZD strategy would be evolutionarily unstable against the Pavlov strategy. To do this, we started with a population of 1000 nodes that are connected via a lattice where each node is connected to 8 other random nodes. Initially, the two strategies were distributed in a random manner so that the ZD strategy would occupy different fractions of the population (0.6 and 0.4) in each simulation run. Then, using the death-birth Moran process, the population was updated over 150,000 time-steps to observe the evolution of the strategies.  

Afterwards, we changed the evolutionary process to the stochastic strategy adoption process used by Santos et al. \cite{santos2006graph}. This enabled us to determine, whether it is the population update process or the network topology that affects the evolutionary stability of the strategies in concern.

\begin{figure}[htbp]
	\begin{center}
		\resizebox{0.5\textwidth}{!}{
			\includegraphics[width=7.0cm, height=5.0cm]{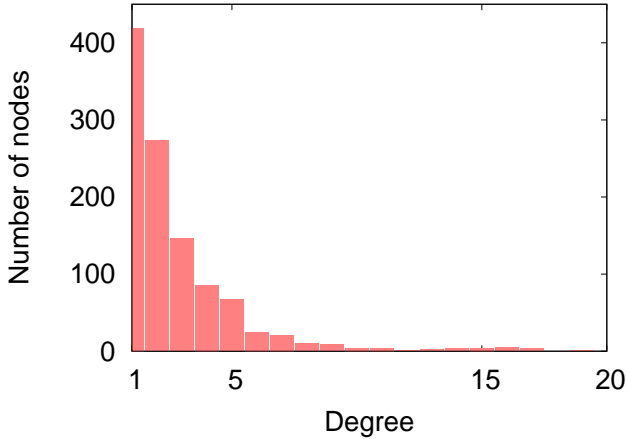} 
		}
		\caption{The degree distribution of the scale-free network used.} \label{sf_dd}
	\end{center}
\end{figure}

Then we replaced the well-mixed population with a non-homogeneous scale-free network with 1000 nodes. As with the case of the previous experiment, the ZD strategy and the Pavlov strategy were assigned randomly among the nodes. With the players spatially distributed in a scale-free network, both the death-birth Moran process and the strategy adoption process were applied separately to observe the effect that they have on the evolution of strategies. The Fig. \ref{sf_dd} depicts the degree distribution of the scale-free network used. As the figure shows, the degree distribution shows a power-law curve that is reminiscent with the typical scale-free networks. 

Next, we changed the initial distribution of the strategies in such a manner that the ZD strategy would occupy the majority of hubs. This was done by sorting the nodes according to their degree and assigning the top 60\% of the nodes with the ZD strategy. In this configuration, the initial average degree of ZD nodes was measured to be 3.4 while the average degree of Pavlov nodes was 1.8. The evolution of strategies was observed under the death-birth Moran process as well as the strategy adoption process. The experiment was repeated with the Pavlov strategy occupying the majority of hubs.

As the next step, we mixed the Pavlov strategy with the general cooperator strategy and the cooperator strategy in separate scale-free networks of players. We tested the evolution of strategies with a random initial distribution of strategies and a strategy distribution where the opponent strategy (GC or cooperator) is initially assigned mostly on hubs. This enabled us to determine whether any observed topological stability of ZD strategy is a unique and inherent property of the stratety itself or whether it is a more general behaviour that could occur with other strategies as well. 

Finally, we observed the evolution of the Pavlov and ZD strategies in non-homogeneous networks while the network heterogeneity was gradually varying. To perform this test, we generated a set of scale-free networks with varying assortativity values by rewiring a scale-free network in a probabilistic manner. Then, both strategies were distributed randomly in each scale-free population in such a manner that the ZD strategy would occupy 60\% of the nodes. Afterwards, the populations were allowed to evolve over 150,000 time-steps under the strategy adoption process and the remaining population fractions of ZD players were recorded. The results were averaged over 40 independent runs.

\section{Results}

\begin{figure}
\begin{center}
\includegraphics[width=7cm, height=5cm]{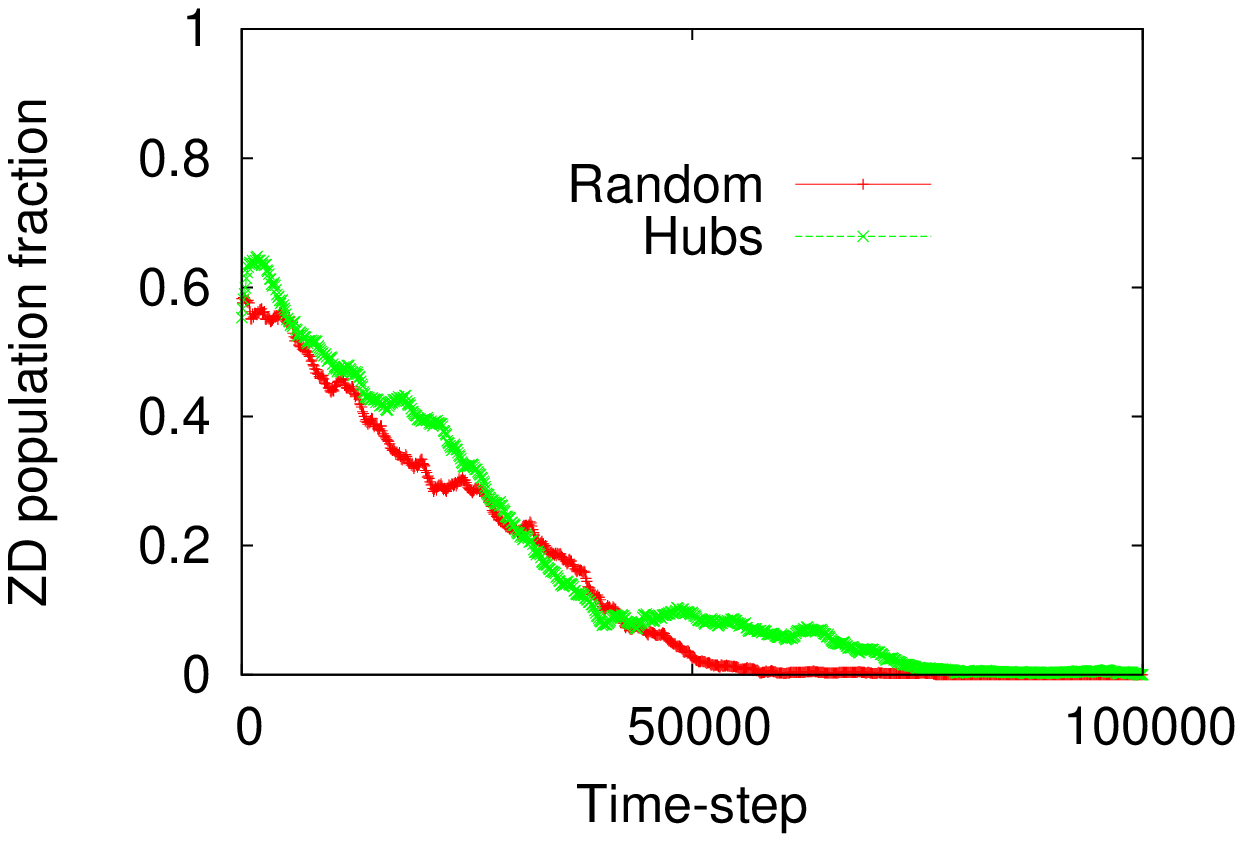}
\caption{The evolution of ZD population fraction against the Pavlov strategy, in scale-free populations of varying initial configurations. The population is allowed to evolve under the death-birth Moran process with 0.1\% replacement rate. In the two initial configurations, ZD strategy is either assigned randomly or assigned more on hubs (in hubs initialisation, the initial average degrees of ZD and Pavlov nodes are 3.4 and 1.8, respectively).}\label{f2}
\end{center}
\end{figure}

\begin{figure}
\begin{center}
\includegraphics[width=7cm, height=5cm]{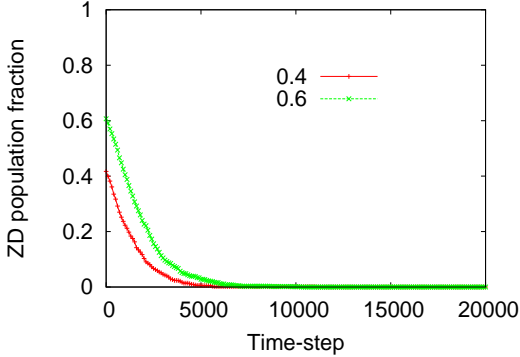}
\caption{The evolution of ZD population fraction against the Pavlov strategy, in a well-mixed population. The population is allowed to evolve under the strategy adoption process. The strategies are initially distributed randomly, with the fraction of ZD nodes being 0.4 and 0.6, respectively.}\label{f3}
\end{center}
\end{figure}

\begin{figure*}[htbp]
\centering
	\begin{subfigure}{0.5\textwidth}
		\centering
		\includegraphics[width=7.0cm, height=5.0cm]{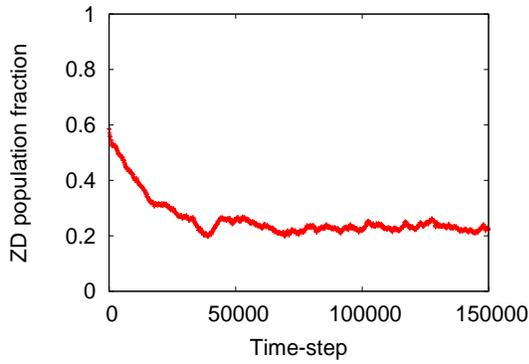}
		\caption{$Randomly-initialised$} 
	\end{subfigure}~
	\begin{subfigure}{0.5\textwidth}
		\centering
		\includegraphics[width=7.0cm, height=5.0cm]{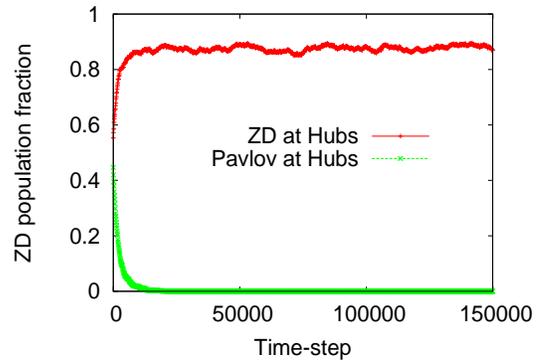}
		\caption{$Hubs-initialised$} 
	\end{subfigure}
\caption{The evolution of ZD population fraction against the Pavlov strategy, in scale-free populations of varying initial configurations. The population is allowed to evolve under the strategy adoption process. (a) ZD and Pavlov strategies initially distributed randomly (b) Majority of hubs are either assigned with ZD or Pavlov (the initial average degrees of hub and non-hub strategies are 3.4 and 1.8, respectively).} \label{f4}  
\end{figure*}

\begin{figure*}[htbp]
\centering
	\begin{subfigure}{0.5\textwidth}
		\centering
		\includegraphics[width=7.0cm, height=5.0cm]{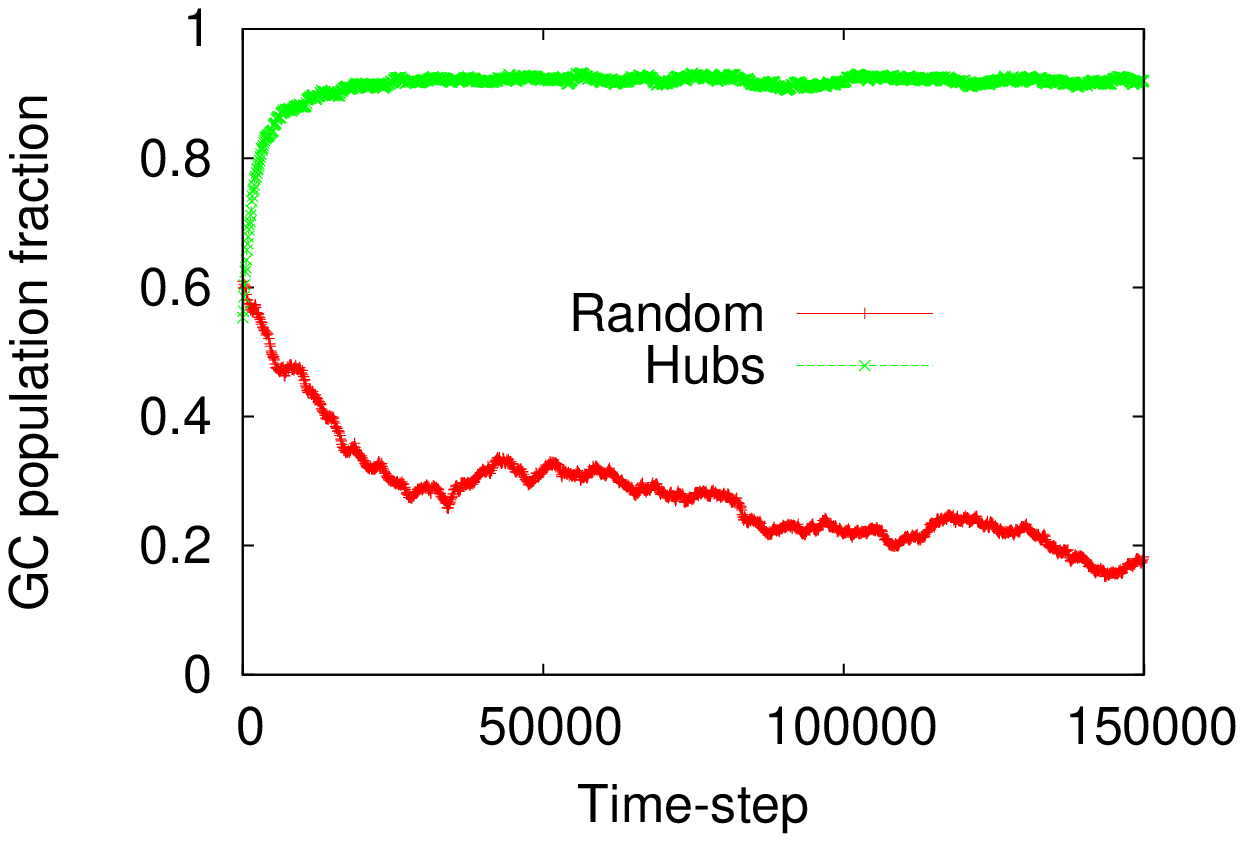}
		\caption{$GC-vs-Pavlov$} 
	\end{subfigure}~
	\begin{subfigure}{0.5\textwidth}
		\centering
		\includegraphics[width=7.0cm, height=5.0cm]{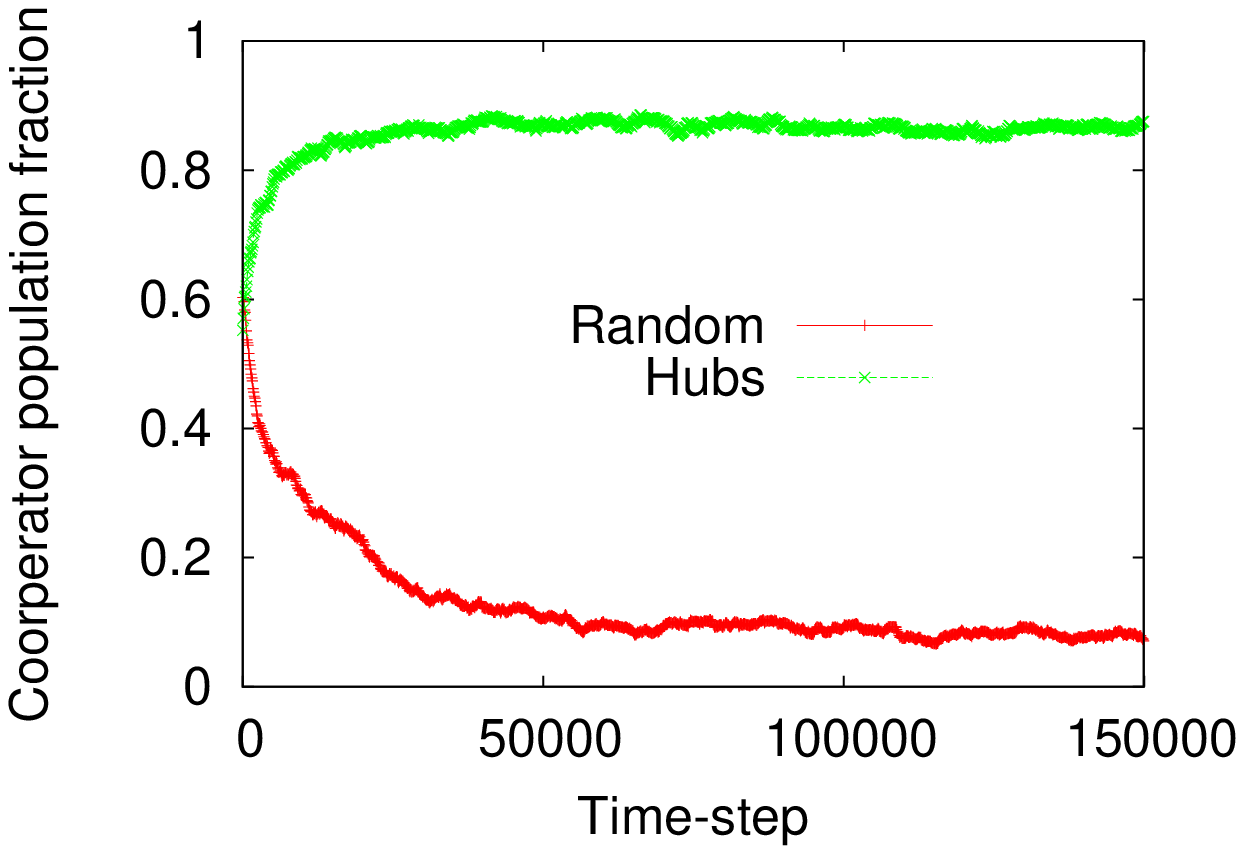}
		\caption{$Cooperator-vs-Pavlov$} 
	\end{subfigure}
\caption{The evolution of GC and cooperator strategies competing against the Pavlov strategy, in scale-free populations of varying initial configurations. The population is allowed to evolve under the strategy adoption process. In the two initial configurations, the competing strategy is either assigned randomly or assigned more on hubs (in hubs initialisation, the initial average degrees of competing nodes and Pavlov nodes are 3.4 and 1.8, respectively).} \label{f5}  
\end{figure*}

\begin{figure*}[htbp]
	\centering
	\begin{subfigure}{0.5\textwidth}
		\centering
		\includegraphics[width=7.0cm, height=5.0cm]{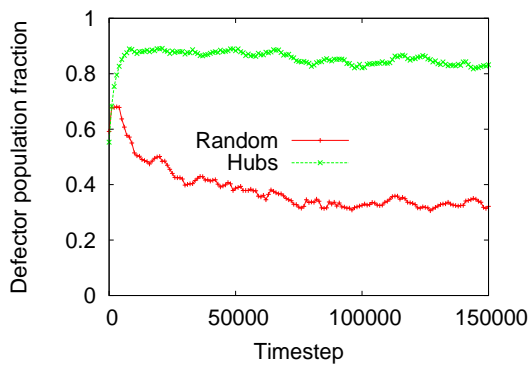}
		\caption{$Defector-vs-Pavlov$} 
	\end{subfigure}~
	\begin{subfigure}{0.5\textwidth}
		\centering
		\includegraphics[width=7.0cm, height=5.0cm]{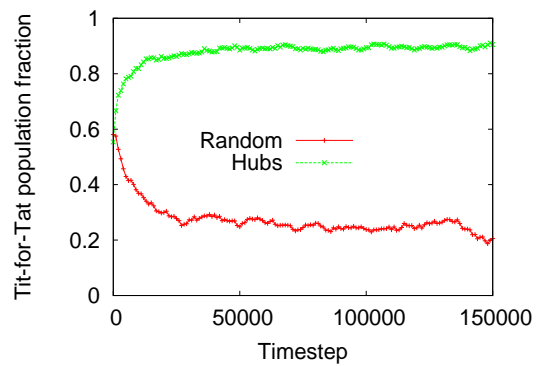}
		\caption{$Tit-for-Tat-vs-Pavlov$} 
	\end{subfigure}
	\caption{The evolution of Defector and Tit-for-tat strategies competing against the Pavlov strategy, in scale-free populations of varying initial configurations. The population is allowed to evolve under the strategy adoption process. In the two initial configurations, the competing strategy is either assigned randomly or assigned more on hubs (in hubs initialisation, the initial average degrees of competing nodes and Pavlov nodes are 3.4 and 1.8, respectively).} \label{f6}  
\end{figure*}

In certain figures in this section, we have limited the number of time-steps shown, when the strategy in concern becomes extinct reasonably quickly. Fig.\ref{f1} shows the evolution of the fraction of ZD nodes when the ZD strategy is mixed with the Pavlov strategy in well-mixed and scale-free populations. The evolutionary process used is the death-birth Moran process. As expected, ZD strategy gradually becomes extinct in a well-mixed population. This confirms that in a homogeneous network, ZD strategy is evolutionarily unstable against the Pavlov strategy that operates as a strong evolutionarily stable strategy, as suggested by Adami and Hintze \cite{adami2013evolutionary}. Moreover, ZD does not survive even in a non-homogeneous population distributed in a scale-free network, when the same evolutionary process is applied.


Next, Fig.\ref{f2} depicts the evolution of the ZD and Pavlov strategies in a scale-free non-homogeneous network of players, under different initial configurations. The figure shows the evolution of the ZD fraction when the strategies are initially distributed randomly as well as when more hubs are assigned with the ZD strategy initially. As the figure depicts, Pavlov clearly dominates and eradicates the ZD strategy, suggesting that irrespective of the initial distribution of strategies, ZD cannot survive when the population is allowed to evolve under the death-birth Moran process.

Fig.\ref{f3} depicts the scenario where a well-mixed population of ZD and Pavlov strategies are allowed to interact with each other over time, according to the strategy adoption process instead of the Moran process. Here too, ZD is gradually eradicated from the population. However, when the same evolutionary process is applied in a scale-free population of players, ZD strategy manages to survive, as shown in Fig\ref{f4}[a]. According to the figure, Pavlov strategy shows weak evolutionary stability, failing to eradicate the ZD strategy completely, when the two strategies are initially assigned randomly. On the other hand, when the ZD is initially assigned to the majority of hubs as depicted in Fig.\ref{f4}[b], ZD manages to become the weak evolutionarily stable strategy over the Pavlov strategy, becoming the dominant strategy in the network. However, as the same figure shows, when the Pavlov strategy is initially assigned to the majority of hubs, it behaves as a strong evolutionarily stable strategy, wiping out the ZD population. This suggests that under the strategy adoption evolutionary process, the evolutionarily unstable ZD strategy may not only survive, but may even become the more prominent strategy in a non-homogeneous population of players.

Then, we mixed the Pavlov strategy with the GC and the cooperator strategies respectively, in a Scale-free population. Fig.\ref{f5} and Fig. \ref{f6} show the evolution of the GC, cooperator, defector and Tit-for-Tat strategy fractions over time, when those strategies are initially placed randomly or mostly on hubs. As the figures depict, the opposing strategies too may survive or dominate the population based on the initial distribution of the strategies, when the population is updated using the strategy adoption evolutionary process. This suggests that topological influence on the evolutionary stability is not limited to the ZD strategy, but may apply to other strategies as well.

\begin{figure}
\begin{center}
\includegraphics[width=7cm, height=5cm]{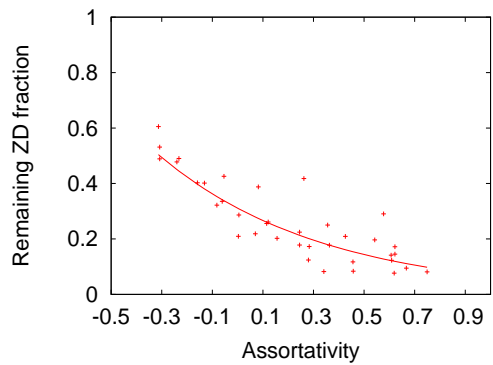}
\caption{The evolution of ZD population fraction against the Pavlov strategy, in scale-free populations with varying network assortativity values. The population is allowed to evolve under the strategy adoption process. The results are averaged over 40 independent runs.}\label{f7}
\end{center}
\end{figure}

Finally, we tested the evolution of the ZD population against the Pavlov strategy while the heterogeneity of the networks are gradually changed. Fig.\ref{f7} shows the variation of the remaining ZD fraction after 150,000 time-steps under the strategy adoption process. As the figure shows, there exits a negative correlation between the network assortativity and the remaining ZD fraction. The actual Pearson correlation value of the two series is -0.85, suggesting a strong negative correlation. Network assortativity is a measure of the similarity or the homogeneity of the mixing patterns of the nodes. Therefore, this result suggests that the effect of network topology on the evolutionary stability of a strategy increases, as the network becomes more heterogeneous.

\section{Discussion}

Network game theory has gained prominence as an interdisciplinary field. Complex networks are increasingly used to model social systems, while the cognitive decision making processes of individuals in such systems is modelled by game theory. It is obvious that the topological connection patterns would influence the decisions made by individual entities, and many questions remain answered in regards to how topology can influence the strategies individual players will choose over time.

In this work, we attempted to evaluate how the network topology of a population of players affect the evolutionary stability of a strategy. In particular, we focused on a class of strategies known as Zero-determinant strategies, which has been demonstrated to be evolutionarily unstable against the Pavlov strategy.

Based on the results gathered from this research, we suggest that network topology has an effect on whether a particular strategy is evolutionarily stable or not. However, the topologically influenced evolutionary stability is a weak evolutionary stability and not a strong evolutionary stability. In other words, the stable strategy would not be able to completely eradicate the competing strategy and the competing strategy would be able to survive within the confines of the network.

We also identified that the topological effect of evolutionary stability is determined by the evolutionary process used. When using the death-birth Moran process to evolve the population, topology does not seem have a significant effect on the evolutionary stability of strategies. However, when the strategy adoption process suggested by Santos et al. \cite{santos2006graph} is applied, topology does have significant effect on evolutionary stability of a strategy within a population. Strategy adoption process takes into account the cumulative payoff of each node in determining whether a strategy should be replaced or not. Therefore, this result suggests that an evolutionarily unstable strategy could survive when they occupy the hubs surrounded by leaf nodes assigned with the evolutionarily stable strategy. In a heterogeneous network of players, hubs tend to have more strategic interactions with their opponents compared to leaf nodes. Thus, a hub with an evolutionarily unstable strategy would continue to be irreplaceable by the neighbouring nodes' strategies, as it would continue to have a higher payoff compared to its immediate neighbours. 

The significance of the evolutionary process may have implications in the real-world networks of strategic players. Moran process would be more appropriate in the biological context where the lifetime of a player is significantly less than the evolutionary time-span. It could be effectively used to model the evolution of species where the strategies are `hard-wired' to the players and the evolution happens through the replacement of players with the replicas of better performing players. However, in the social context, the evolution of strategies may be driven by the adoption of strategies by the players based on the performance of their neighbouring players. In other words, a stochastic strategy adoption process could be used to model the evolution of strategies when the lifetime of a player maybe considerably larger than the time-span of evolution. Examples of such situations involve the interactions that happen in corporate sector and financial markets. There, it is often observable that the players continually adopt the strategies of other players, in their struggle to survive. Thus, the strategy adoption evolutionary update process may be more relevant when the evolution of strategies is applied in the social context. Accordingly, topological effect on the evolutionary stability of strategies may be more prevalent in the social context, compared to the biological context.

Further, it is important to note that not only the topology, but also the initial distribution of the strategies within the network too plays a significant role in shaping the evolution of the strategies. For instance, when the evolutionarily unstable strategy occupies hubs as opposed to the leaf nodes at the initiation of the evolution, it even manages to become the more prominent strategy within the network over time, resembling a weak evolutionarily stable strategy. 

Even though we mainly focused on the ZD and Pavlov strategies in this work, we could replicate similar observations with other well-known strategies such as the general cooperator and cooperator strategies, competing against the Pavlov strategy. This could mean that the variation of evolutionary stability due to topological stability of strategies is a more general phenomena that may be applicable to most strategies that are competing with each other. In addition, the topological effect on the evolutionary stability may be observed for other games other than the Prisoner's dilemma game. For instance, it would be interesting to observe how the extensive form games, where the temporal dimension is built-in to the strategic decisions, are affected by the topological placement of the players. Such studies may be useful in accurately predicting the evolutionary stability of strategies in scenarios such as auctions, which can be modelled with extensive form games. 

In conclusion, we could identify the following three factors that determine the topological stability of strategies in a non-homogeneous network. Network topology, evolutionary process and the initial distribution of the strategies. By varying these three factors, an evolutionarily unstable strategy may be able to survive and may even operate as a weak evolutionarily stable strategy, in a population of players connected in a non-homogeneous topology. Based on our observations, the topological stability of strategies may be more prevalent in the social context of the evolution of strategies, in comparison to the biological context.

\section*{Acknowledgment}

The authors would like to thank  Michael Harr\`e of the Centre for Complex Systems Research at the University of Sydney and  Chistoph Adami and  Arend Hintze of Michigan State University for their valuable comments and suggestions.

\bibliography{topo}
\bibliographystyle{ieeetran}

\end{document}